\documentclass[conference]{IEEEtran}
\IEEEoverridecommandlockouts
\usepackage{cite}
\usepackage{mathrsfs}
\usepackage{amsmath,amssymb,amsfonts}
\usepackage{algorithmic}
\usepackage{graphicx}
\usepackage{textcomp}
\usepackage{xcolor}
\usepackage{multirow}
\usepackage{booktabs}
\usepackage{bbding}
\def\BibTeX{{\rm B\kern-.05em{\sc i\kern-.025em b}\kern-.08em
    T\kern-.1667em\lower.7ex\hbox{E}\kern-.125emX}}

\title{BeatFM: Improving Beat Tracking with Pre-trained Music Foundation Model}

\author{
\IEEEauthorblockN{
Ganghui Ru\textsuperscript{1, $\star$}
Jieying Wang\textsuperscript{2, $\star$},
Jiahao Zhao\textsuperscript{3},
Yulun Wu\textsuperscript{1},
Yi Yu\textsuperscript{4},
Nannan Jiang\textsuperscript{2, $\dag$},
Wei Wang\textsuperscript{2, $\dag$},
and Wei Li\textsuperscript{1, 5, $\dag$} 
}

\IEEEauthorblockA{
\textsuperscript{1}School of Computer Science, Fudan University, Shanghai, China \\
\textsuperscript{2}Naval Medical Center, PLA, China \\
\textsuperscript{3}Graduate School of Informatics, Kyoto University, Kyoto, Japan \\
\textsuperscript{4}Graduate School of Advanced Science and Engineering, Hiroshima University, Hiroshima, Japan \\
\textsuperscript{5}Shanghai Key Laboratory of Intelligent Information Processing, Fudan University, Shanghai, China
}

\IEEEauthorblockA{
ghru23@m.fudan.edu.cn, 
\{20110240018, wwang\_fd, weili-fudan\}@fudan.edu.cn, 
\{wanwan0426, jiangnannannavy\}@163.com, 
 \\
zhao.jiahao.56h@st.kyoto-u.ac.jp, 
yiyu@hiroshima-u.ac.jp 
}
}

\begin{document}

\maketitle
\footnotetext{$\star$These authors contributed equally to this work}
\footnotetext{$\dag$Corresponding author}

\begin{abstract}
Beat tracking is a widely researched topic in music information retrieval. 
However, current beat tracking methods face challenges due to the scarcity of labeled data, which limits their ability to generalize across diverse musical styles and
accurately capture complex rhythmic structures.
To overcome these challenges, we propose a novel beat tracking paradigm BeatFM, which introduces a pre-trained music foundation model and leverages its rich semantic knowledge to improve beat tracking performance.
Pre-training on diverse music datasets endows music foundation models with a robust understanding of music, thereby effectively addressing these challenges.
To further adapt it for beat tracking, we design a plug-and-play multi-dimensional semantic aggregation module, which is composed of three parallel sub-modules, each focusing on semantic aggregation in the temporal, frequency, and channel domains,  respectively.
Extensive experiments demonstrate that our method achieves state-of-the-art performance in beat and downbeat tracking across multiple benchmark datasets.
\end{abstract}


\section{Introduction}
\label{sec:intro}

With the rapid growth of the digital music industry, automatically identifying beat sequences in musical works has become an urgent challenge in Music Information Retrieval (MIR). 
The goal of beat tracking is to detect the temporal positions of beats within a music signal, which is essential for understanding its rhythmic structure and dynamic variations. 
Furthermore, accurate beat tracking serves as a crucial component for various downstream music information retrieval (MIR) tasks, such as music transcription \cite{transcription1} and chord estimation \cite{chord1}, among others \cite{ru2023improving, duan2023melody}.

Despite its wide-ranging applications, beat tracking remains a challenging  problem due to the varied and intricate nature of musical rhythms across different genres. 
Music exhibits a vast diversity of rhythmic patterns, ranging from steady and regular beats, as commonly found in electronic dance music, to complex syncopations and irregular tempos often present in classical compositions.
Additionally, factors such as polyrhythms and unexpected changes in rhythmic structure make it difficult to reliably detect beats across compositions.

Current beat tracking methods \cite{bock2020, tftrans, beattrans, beatthis} have made notable progress in tackling these challenges by developing innovative algorithms and more sophisticated models. However, they are fundamentally constrained by several key limitations. First, the scarcity of labeled data for training, particularly in niche genres or complex rhythmic patterns, remains a significant challenge. Second, the imbalance in music genres within training data further exacerbates this issue, hindering the performance of current systems across diverse musical styles. As a result, the performance reaches a ceiling, preventing them from achieving robust and accurate beat tracking.

To overcome these limitations, we propose a novel beat tracking paradigm that departs from traditional approaches of designing task-specific feature extraction models. Specifically, we introduce a pre-trained music foundation model, which has been trained on extensive and diverse music datasets. The foundation model provides a deeper, more robust understanding of musical structures, enabling the capture of complex rhythmic patterns, dynamic tempo variations, and diverse musical styles. By leveraging the rich semantic knowledge embedded in the pre-trained music foundation model, our approach circumvents the limitations imposed by the scarcity of labeled data and genre imbalance. 

Additionally, to fully unlock the potential of the pre-trained model for beat tracking, we design a plug-and-play multi-dimensional semantic aggregation module to further enhance the semantic representation of music. This module consists of three parallel sub-modules, each designed to capture a distinct aspect of musical information. The temporal sub-module focuses on identifying rhythmic patterns and modeling the sequential dynamics of the music, enabling the model to capture the flow and structure of beats. The frequency sub-module captures harmonic elements, enhancing the model's understanding of musical structure and tonal relationships. The channel sub-module aggregates information from different semantic levels, allowing the model to adaptively focus on both detailed musical textures and overarching structural patterns. The synergy among these semantic aggregation sub-modules enables a more robust representation of musical data, significantly enhancing beat and downbeat tracking precision. We conducted extensive experiments across multiple benchmark datasets, demonstrating that our approach achieves state-of-the-art performance in both beat and downbeat tracking. In summary, our contributions are as follows:
\begin{itemize}
  \item We propose a novel beat tracking paradigm, BeatFM, which integrates a pre-trained music foundation model to address key challenges in traditional data-driven methods, such as data scarcity and genre imbalance. 
  \item We design a plug-and-play multi-dimensional semantic aggregation module that enhances the semantic representation of music by effectively integrating information across temporal, frequency, and channel domains.
  \item Extensive experiments on multiple benchmark datasets show that our approach achieves state-of-the-art performance in both beat and downbeat tracking.
\end{itemize}

\section{Related Work}
\label{sec:format}
Beat tracking is a fundamental task in music information retrieval, aimed at detecting the rhythmic structure of music. Early beat tracking systems \cite{Goto_2001, Dixon_2001} were often based on signal processing techniques, such as onset detection, autocorrelation, and phase-based methods. 
With the rise of deep learning, models like Recurrent Neural Networks (RNNs) \cite{rnn1, rnn2}, Convolutional Neural Networks (CNNs) \cite{cnn1, ubeat}, and Temporal Convolutional Networks (TCNs) \cite{tcn1, bock2020} have been widely adopted to capture temporal dependencies and rhythmic patterns. Transformer-based models have recently gained attention for their ability to capture long-range dependencies more effectively. Huang \textit{et al.} \cite{tftrans} proposed a Transformer-based architecture for beat tracking, significantly improving model accuracy and performance in capturing rhythmic structures. 
In subsequent works, Zhao \textit{et al.} \cite{beattrans} employed dilated self-attention mechanisms combined with music separation techniques to better capture rhythmic variations and long-range dependencies, improving the model's ability to handle diverse musical styles.
Similarly, Cheng \textit{et al.} \cite{lhtrans} introduced a two-stage approach involving a low-resolution encoder and a high-resolution decoder. The low-resolution encoder is effective in capturing global features with more balanced data, while the high-resolution decoder focuses on fine-grained beat prediction, allowing for better temporal resolution in tracking beats.

In addition, self-supervised learning has recently been applied to beat tracking. For example, Heydari \textit{et al.} \cite{singingbeattracking} explored the use of pre-trained self-supervised speech models as feature extractors for beat tracking in vocal performances. Their approach specifically focuses on the application of speech representation models to capture rhythmic features in singing, which is tailored to vocal music.  Other works, such as Desblancs \textit{et al.} \cite{zeronotesamba}, have proposed unsupervised beat tracking methods, aiming to eliminate the dependency on labeled data. In contrast, Chiu \textit{et al.} \cite{chiu2023local} and Foscarin \textit{et al.} \cite{beatthis} proposed alternatives to traditional post-processing techniques to refine beat predictions more efficiently.

Despite the progress made in deep learning-based beat tracking, existing methods still face fundamental limitations. The primary challenges, such as data scarcity and genre imbalance, remain largely unaddressed. To overcome these issues, we propose a novel beat tracking paradigm, BeatFM, which integrates a pre-trained music foundation model and leverages its robust semantic knowledge. Unlike traditional approaches that design specialized network structures for feature extraction, our approach focuses on exploring the potential of music foundation models for downstream beat tracking tasks.

\section{Method}
\label{sec:format}
\begin{figure*}[ht]
	\centering
	\includegraphics[width=0.95\linewidth]{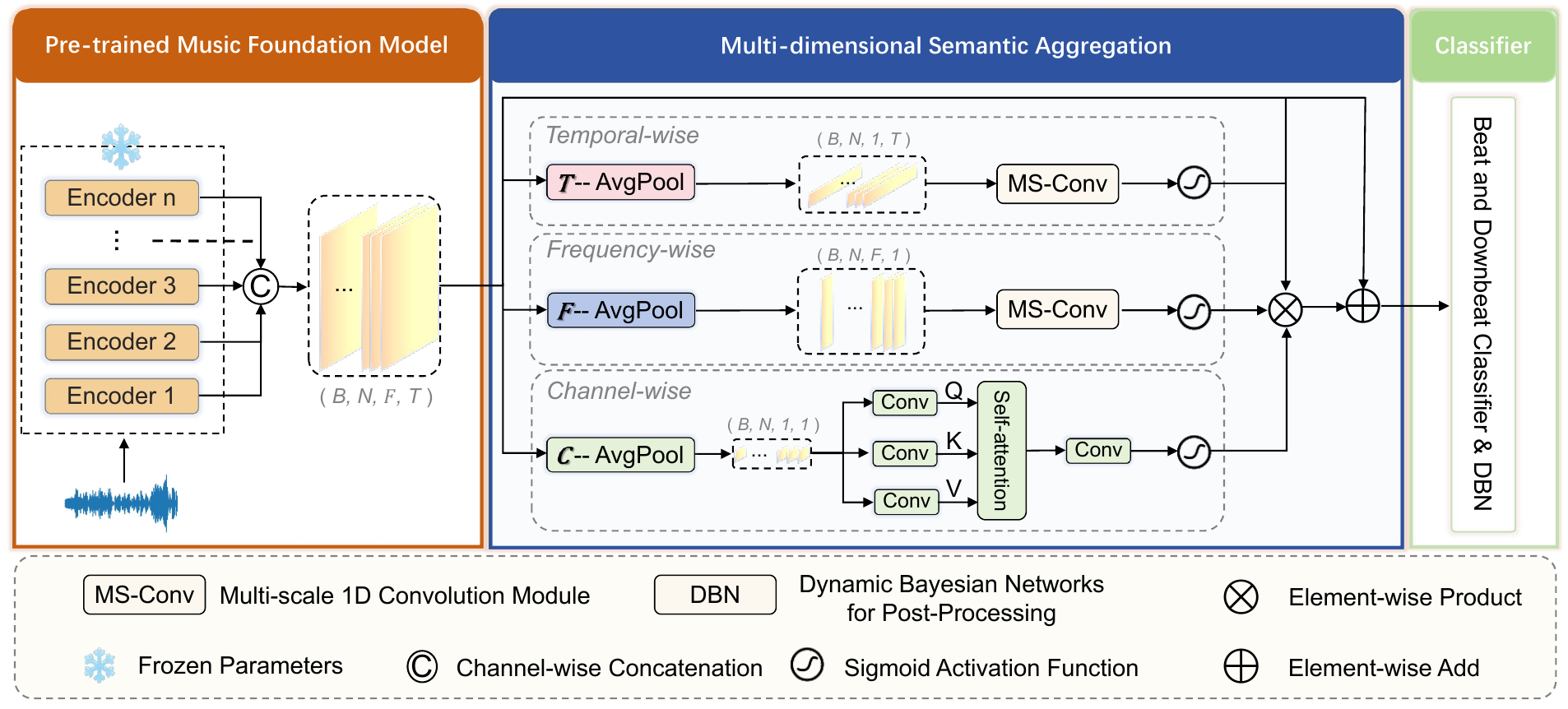}	
	\caption{Overview of our proposed BeatFM, consisting of Pre-trained Foundation Model, Multi-dimensional Semantic Aggregation Module and Classifier.}
	\label{figure1}
\end{figure*}

As illustrated in Figure \ref{figure1}, our proposed BeatFM consists of three parts: the pre-trained music foundation model, the multi-dimensional semantic aggregation module, and the beat and downbeat classifier. 
\subsection{Pre-trained Music Foundation Model}
Pre-trained music foundation models have shown great promise in improving the performance of downstream MIR tasks, particularly those requiring the understanding of complex musical structures \cite{mertech}. Traditional deep learning-based methods often face challenges such as data scarcity and genre imbalance.  In contrast, pre-trained music foundation models, which are trained on large-scale and diverse music datasets, offer a solution by leveraging their ability to generalize across various musical genres and structures. This enables them to perform effectively in downstream tasks, even when faced with limited or imbalanced data.  Recent works, such as MERT \cite{mert} and MusicFM \cite{musicfm}, have highlighted the potential of leveraging pre-trained models to enhance performance in various MIR tasks by providing a richer semantic representation of music.

The core architecture of the music foundation models consists of multiple stacked Transformer encoders, where each encoder is designed to capture different levels of musical semantics present in the audio signal. These encoders process the input audio features sequentially, learning hierarchical representations that range from low-level acoustic details to high-level musical patterns. 

Given input audio x, we obtain its corresponding multi-level semantic feature representation. We then concatenate these features to form a unified representation $h \in \mathbb{R}^{b*n*f*t}$:
\begin{equation}
\begin{aligned}
[h_1, h_2, ..., h_n] = FM(x)
\end{aligned}
\end{equation}
\begin{equation}
\begin{aligned}
h = Concat([h_1, h_2, ..., h_n])
\end{aligned}
\end{equation}
where $FM$ represents the pre-trained music foundation model.

\subsection{Multi-dimensional Semantic Aggregation Module}
The multi-dimensional semantic aggregation module (MSAM)  is designed to enhance the performance of the pre-trained music foundation model for beat tracking tasks. Although the foundation model provides a strong semantic representation of music, it is not specifically tailored to the unique demands of beat tracking. Therefore, the goal of MSAM is to enhance the pre-trained model's ability to capture rhythmic patterns and temporal dynamics, which are critical for accurate beat detection. The module achieves this through three carefully designed sub-modules, which aggregate information across temporal, frequency, and channel dimensions.

\textbf{Temporal-wise Aggregation.} Research \cite{sensitivity2} in music cognition science indicates that the human brain is highly sensitive to multiple levels of temporal periodicity in complex auditory sequences. This sensitivity further reveals the intrinsic reason for employing hierarchical structures in music composition. To model these temporal hierarchies, we propose the temporal-wise aggregation sub-module. 
To emphasize the temporal variations in the input feature map $h$, we first apply temporal-wise average pooling \textit{T-AvgPool} to obtain $h_t \in \mathbb{R}^{b*n*t}$. This operation reduces the computational complexity of the sub-module while preserving temporal structures critical for capturing rhythmic patterns.
Subsequently, $h_t$ is processed by a multi-scale 1D convolution module (MS-Conv), which consists of $M$ parallel 1D convolutions with varying dilation rates. The outputs of these convolutions are then concatenated and passed through an MLP layer that maps the resulting features back to the original dimensional space. Finally, temporal attention $Attn_t \in \mathbb{R}^{b*n*t}$ is generated using a Sigmoid activation function, which selectively emphasizes and suppresses specific temporal regions. The entire process can be formulated as:
\begin{equation}
\mathbf{h}_t = \textit{T-AvgPool}(\mathbf{h})
\end{equation}
\begin{equation}
\mathbf{h}_t^{(i)} = Conv1d_{dilated(r_i)}(\mathbf{h}_t), \quad \forall i \in \{1, 2, \dots, M\}
\end{equation}
\begin{equation}
Attn_t = \sigma(MLP(Concat[\mathbf{h}_t^{(1)}, \mathbf{h}_t^{(2)}, ..., \mathbf{h}_t^{(M)}]))
\end{equation} where $\sigma(\cdot)$ denotes the Sigmoid activation function. The temporal-wise sub-module allows the model to effectively learn rhythmic patterns over different time scales, providing a more robust feature representation for beat tracking task.


\textbf{Frequency-wise Aggregation.} The frequency-wise aggregation sub-module focuses on capturing harmonic and melodic features across different frequency bands. In music theory, harmonic and melodic changes are more likely to occur at beat positions than at non-beat positions, which provides essential cues for rhythm perception \cite{Dixon_2001}. This insight forms the basis for the design of frequency-wise aggregation sub-module. Hence, we introduce the frequency-wise aggregation sub-module to effectively model these frequency-based variations. 
The frequency-wise aggregation sub-module begins with frequency-wise average pooling \textit{F-AvgPool} to obtain $h_f \in \mathbb{R}^{b*n*f}$, after which it follows the same structure as the temporal-wise aggregation sub-module to compute the frequency attention $Attn_f \in \mathbb{R}^{b*n*f}$. However, the parameters are not shared due to the need to model different aspects of the music signal.


\textbf{Channel-wise Aggregation.} The channel-wise aggregation sub-module is designed to integrate and aggregate multi-level semantic features extracted from different encoding layers of the music foundation model. 
This module focuses on facilitating information exchange and collaboration between channels, capturing the rich and diverse information present at different feature levels. 
In this sub-module, we first apply channel-wise average pooling \textit{C-AvgPool} to obtain $h_c \in \mathbb{R}^{b*n}$. Then, we utilize the self-attention mechanism to  compute $Attn_c \in \mathbb{R}^{b*n}$, which dynamically weighs the importance of each channel. The implementation details are as follows:
\begin{equation}
\mathbf{h}_c = \textit{C-AvgPool}(\mathbf{h})
\end{equation}
\begin{equation}
Q = Conv(h_c), K = Conv(h_c), V = Conv(h_c)
\end{equation}
\begin{equation}
\begin{aligned}
Attn_c &= \sigma(Conv(\textit{self-attn}(Q, K, V)) \\
       &= \sigma(Conv(Softmax(\frac{QK^T}{\sqrt{C}})V))
\end{aligned}
\end{equation}

To integrate the outputs from the three sub-modules, we unify them into a single attention map, which is then applied to the input feature map $h$:

\begin{equation}
Attn = Attn_t \cdot Attn_f \cdot Attn_c
\end{equation}
\begin{equation}
\tilde{h} = h + Attn * h
\end{equation}

By focusing on temporal, frequency, and channel aspects, we enable the model to capture both fine-grained rhythmic details and high-level musical structures. The integration of multi-dimensional information allows the model to effectively handle complex music with varying tempos and genres, providing a more robust feature representation for the downstream beat tracking task.

\subsection{Beat and Downbeat Classifier}
The beat and downbeat classifier is the final component of our model. It takes the features aggregated by the multi-dimensional semantic aggregation module and converts them into beat and downbeat probabilities using fully connected layers. These probabilities are then refined through a Dynamic Bayesian Network \cite{dbn} to ensure temporal consistency and improve detection reliability, which is a common post-processing technique employed by many existing beat tracking models.

\section{experiments}
\subsection{Datasets and Metrics}
Following previous work \cite{ubeat, lhtrans}, we utilize seven widely used music datasets with beat annotations to train, validate, and test the proposed model. These datasets cover a diverse range of musical styles and rhythms, ensuring a comprehensive evaluation of the model's performance across different scenarios. Specifically, \textit{Beatles} \cite{beatles}, \textit{RWC Popular} \cite{rwc_popular}, and \textit{Harmonix} \cite{harmonix} datasets are used for training. \textit{Ballroom} \cite{ballroom1, ballroom2}, \textit{Hainsworth} \cite{hainsworth}, and \textit{SMC} \cite{smc} datasets are used for both training and testing in an 8-fold cross-validation manner. \textit{GTZAN} \cite{gtzan2, gtzan1} dataset is exclusively used for testing, providing an additional benchmark to assess the generalization capability of the model. 

For the evaluation of beat and downbeat tracking, we adopt three widely recognized metrics: F1-measure, CMLt (Correct Metric Level), and AMLt (Allowed Metric Level with tolerance for off-beat, double, or half-tempo). The latter two metrics primarily assess the proportion of correctly predicted beat sequences that align with the ground truth, reflecting the model's ability to consistently predict rhythmic patterns over time. Following \cite{beatles}, we apply a tolerance window of 70 ms for beat alignment, which is standard in beat tracking research.  The \textit{mir\_eval} library \cite{raffel2014mir_eval} is used to calculate all metrics, which ensures a thorough evaluation of beat tracking accuracy.

\begin{table*}[h]
  \centering
  \caption{Comparison with other state-of-the-art beat tracking models and two baseline models on the \textit{GTZAN} dataset.}
  \fontsize{5}{5}\selectfont
  \resizebox{\textwidth}{!}{
  
    \begin{tabular}{cccccccc}
    \toprule
          &       & \multicolumn{3}{c}{\textbf{Beat Accuracy}} & \multicolumn{3}{c}{\textbf{Downbeat Accuracy}} \\
    \cmidrule{3-8}    
    \textbf{Dataset} & \textbf{Model} & \textbf{F-Measure} & \textbf{CMLt} & \textbf{AMLt} & \textbf{F-Measure} & \textbf{CMLt} & \textbf{AMLt} \\
    \midrule
    \multirow{7}{*}{GTZAN} 
          & TCN \cite{bock2020} & 88.5 & 81.3 & 93.1 & 67.2 & 64.0 & 83.2 \\
          & Beat trans \cite{beattrans} & 88.5 & 80.0 & 92.2 & 71.4 & 66.5 & 84.4 \\
          & Beat This \cite{beatthis} & 88.9 & 79.9 & 89.4 & 75.5 & 60.8 & 75.5 \\ 
          & MERT \cite{mert} & 87.3 & 78.4 & 90.7 &  74.8 & 69.3 & 86.1 \\
          & MusicFM \cite{musicfm} & 86.1 & - & - & 78.5 & - & - \\
          \cmidrule{2-8}
          & \textbf{BeatFM (MERT)}  & \textbf{89.5} & \textbf{82.7} & \textbf{93.9} & 76.7 & 70.2 & 85.4 \\
          & \textbf{BeatFM (MusicFM)}  & 89.1 & 80.6 & 93.5 & \textbf{79.6} & \textbf{74.4} & \textbf{88.7} \\
    \bottomrule
    \end{tabular}%
    }
  \label{testing_results}%
\end{table*}%

\subsection{Implementation Details}

Our model employs a multi-task learning framework, jointly predicting beat and downbeat tracking. We utilize binary cross-entropy loss to supervise the training process. Following \cite{bock2020}, we apply the same label broadening strategy for both beat and downbeat annotations. Specifically, frames adjacent to the annotated beat frames (\(\pm2\) frames) are also labeled as beats but with reduced weights of 0.5 and 0.25, respectively.

In both the temporal-wise and frequency-wise aggregation sub-modules, the multi-scale 1D convolution modules (MS-Conv) consist of \( M=4 \) parallel 1D convolutions, with dilation rates set to \([1, 2, 4, 8]\). This design captures patterns across multiple temporal or frequency scales, effectively modeling complex rhythmic structures and harmonic variations.

For the training and validation sets, all music segments are divided into 15-second clips with a 5-second overlap. To prevent any overlap, clips from the same music piece are assigned exclusively to either the training set or the validation set. Evaluation on the test set is performed on complete, unsegmented music pieces. The model is trained using the Adam optimizer with a learning rate of \(3 \times 10^{-4}\) and a batch size of 16. Early stopping is applied when the validation loss does not decrease for 20 epochs.

\subsection{Baseline and SOTA Models}
We compare our proposed BeatFM with several SOTA models, including TCN \cite{bock2020}, Beat trans \cite{beattrans}, and Beat This \cite{beatthis}. Additionally, we establish two baseline models based on pre-trained music foundation models: MERT \cite{mert} and MusicFM \cite{musicfm}, which only connect a classifier along with DBN post-processing on top of the frozen pre-trained music foundation models. 
\begin{itemize}
  \item \textbf{MERT} employs a Residual Vector Quantization-Variational Autoencoder (RVQ-VAE) to capture the acoustic properties and uses a Constant-Q Transform (CQT) as a musical teacher model to guide the learning of music-specific features, making it suitable for local frame-level sequence labeling tasks.
  \item \textbf{MusicFM} is another pre-trained music foundation model, inspired by the speech recognition model BEST-RQ \cite{bestrq}. It adopts a random projection quantizer during the tokenization phase, enabling efficient modeling of long-term contextual dependencies in musical data.
\end{itemize}  

To ensure a fair comparison, we reproduced MERT's results on the \textit{GTZAN} dataset, as its originally reported results did not strictly follow beat tracking community conventions. Similarly, for the Beat This model, which was trained with additional data, we only report its results on the \textit{GTZAN} dataset using the standard training setup.

\section{Results}

\begin{table}[ht]
    \centering
    \caption{Comparison with other beat tracking models on datasets used in an 8-fold cross-validation setup.}
    \label{table:beat_accuracy}
    \begin{tabular}{cccc}
        \toprule
        \textbf{Dataset} & \textbf{Model} & \textbf{Beat F1} & \textbf{Downbeat F1} \\
        \midrule
        \multirow{6}{*}{Ballroom}
        & TCN \cite{bock2020} & 96.2 & 91.6  \\
        & Beat trans \cite{beattrans} & 96.8 & 94.1  \\
        & MERT \cite{mert}  & 95.7 &  93.2 \\
        & MusicFM \cite{musicfm} & 95.1 & 94.3  \\
        & \textbf{BeatFM (MERT)}  & \textbf{97.0} & 94.3  \\
        & \textbf{BeatFM (MusicFM)}  & 96.7 & \textbf{95.2}  \\
        \midrule
        \multirow{6}{*}{Hainsworth}
        & TCN \cite{bock2020} & 90.4 & 72.2  \\
        & Beat trans \cite{beattrans} & 90.2 & 74.8  \\
        & MERT \cite{mert}  & 89.6 & 74.5  \\
        & MusicFM \cite{musicfm} & 89.2 & 75.7  \\
        & \textbf{BeatFM (MERT)}  & \textbf{90.9} & 75.8  \\
        & \textbf{BeatFM (MusicFM)}  & 90.5 & \textbf{77.6}  \\
        \midrule
        \multirow{6}{*}{SMC}
        & TCN \cite{bock2020} & 55.2  & - \\
        & Beat trans \cite{beattrans} & 59.6 & -  \\
        & MERT \cite{mert}  & 60.1 & -  \\
        & MusicFM \cite{musicfm} & 59.2 &  - \\
        & \textbf{BeatFM (MERT)}  & \textbf{61.3} & -  \\
        & \textbf{BeatFM (MusicFM)}  & 60.5 & -  \\
        \bottomrule
    \end{tabular}
    \label{beat_accuracy}
\end{table}

\subsection{Comparison with Previous Works}

Table \ref{testing_results} summarizes the performance of our proposed model, BeatFM, compared with SOTA beat tracking models and two baseline models on the \textit{GTZAN} dataset.  

Compared to the previous best-performing model, Beat This \cite{beatthis}, our BeatFM model achieves superior performance in both beat and downbeat tracking. Specifically, BeatFM achieves an F1-measure of 79.6\% in downbeat tracking, marking a significant improvement of 4.1\% over Beat This, which achieved 75.5\%.  In comparison to baseline models (MERT and MusicFM), our BeatFM variants consistently show robust performance improvements. For beat tracking, BeatFM (MERT) improves the F1-measure from 87.3\% to 89.5\%, while BeatFM (MusicFM) improves it from 86.1\% to 89.1\%. Similarly, in downbeat tracking, BeatFM (MusicFM) achieves an F1-measure of 79.6\%, compared to 78.5\% for the baseline MusicFM model. These results demonstrate the stability and generalizability of our proposed model across different pre-trained foundation models.

Table \ref{table:beat_accuracy} presents the performance of our model on the Ballroom, Hainsworth, and \textit{SMC} datasets using an 8-fold cross-validation setup. Across these datasets, BeatFM consistently outperforms previous SOTA models and baseline models. The \textit{SMC} dataset, composed of Western classical music with complex and irregular rhythms, is particularly challenging for beat tracking. On this dataset, BeatFM exhibits clear advantages over existing models, highlighting its ability to handle rhythmically intricate and diverse musical compositions.

\begin{table}[ht]
    \centering
    \caption{Ablation study on beat tracking with the three semantic aggregation sub-modules on the \textit{GTZAN} dataset.}
    \label{abation}
    \begin{tabular}{ccc|c}
        \toprule
        Temporal & Frequency & Channel  & F-measure\\
        \midrule
         &  & & 87.3 \\
        \Checkmark & & & 88.2 \\
         & \Checkmark& & 87.9 \\
         & & \Checkmark  & 87.7 \\
        \Checkmark & \Checkmark &  & 88.9 \\
        \Checkmark & &\Checkmark  & 88.6 \\
         &\Checkmark & \Checkmark  & 88.3 \\
        \Checkmark & \Checkmark & \Checkmark  & 89.5 \\
        \bottomrule
    \end{tabular}
\end{table}

Overall, these results demonstrate that BeatFM is a robust and generalizable framework for beat and downbeat tracking tasks. By leveraging the strengths of pre-trained foundation models and task-specific semantic aggregation, our approach consistently surpasses both SOTA methods and baseline models across a variety of datasets.

\subsection{Ablation Study}
To investigate the contributions of each semantic aggregation sub-module in our proposed model, we conduct ablation studies on \textit{GTZAN} dataset, as summarized in Table \ref{abation}. 
These studies aim to evaluate the independent contributions of the temporal-wise, frequency-wise, and channel-wise sub-modules, as well as their combined effects on the overall model performance.
The baseline model, which excludes all sub-modules, achieves the lowest performance, with an F-measure of only 87.3\%.  
When incorporating any single sub-module into the baseline model, we observe significant improvements in F-measure, confirming that each sub-module independently contributes positively to the model’s effectiveness.
Furthermore, combining any two sub-modules yields better results than using a single sub-module alone, suggesting that the sub-modules complement each other and work together effectively. Across  all ablation experiments, the fully integrated model achieves the highest F-measure of 89.5\%. This result highlights the synergistic effect of the multi-dimensional semantic aggregation module, where the sub-modules work together to capture a comprehensive representation of musical signals.

\section{conclusion}
In this paper, we proposed BeatFM, a novel beat tracking paradigm that integrates pre-trained music foundation models with a multi-dimensional semantic aggregation module. By leveraging the rich semantic representations of pre-trained models and tailoring the aggregation module to the specific characteristics of beat tracking tasks, our approach addresses key challenges such as data scarcity and genre imbalance. Comprehensive experiments on multiple benchmark datasets demonstrate that BeatFM outperforms existing state-of-the-art models in both beat and downbeat tracking tasks. Our work demonstrates the potential of leveraging pre-trained music foundation models for beat tracking tasks and provides a solid foundation for future research. In the future, we plan to further explore lightweight adaptations of foundation models for real-time beat tracking and investigate their applicability to more diverse musical styles and complex rhythmic structures. 

\section{Acknowledgement}
This work was supported by NSFC(62171138).

\bibliographystyle{IEEEbib}
\bibliography{icme2025references}

\end{document}